\documentclass[superscriptaddress,prl,twocolumn,amsmath,amssymb,floatfix, showkeys]{revtex4}
\usepackage{graphicx}
\usepackage{bm}
\usepackage{amssymb}
\usepackage{color}
\usepackage{epsfig}
\usepackage{subfigure}
\usepackage{graphicx}
\usepackage{epsfig}
\usepackage{subfigure}
\usepackage{amssymb,amsfonts,amsmath}

\newcommand{\be}{\begin{equation}}
\newcommand{\ee}{\end{equation}}
\begin{document}

\title {On relaxations and aging of various glasses}


\author{Ariel Amir}
\affiliation { Department of Condensed Matter Physics, Weizmann Institute of Science, Rehovot, 76100, Israel}
\affiliation { Department of Physics, Harvard University, Cambridge, MA 02138, USA}
\author{Yuval Oreg}
\author{Yoseph Imry}
\affiliation { Department of Condensed Matter Physics, Weizmann Institute of Science, Rehovot, 76100, Israel}

\begin{abstract}
Slow relaxation occurs in many physical and biological systems. `Creep' is an
example from everyday life: when stretching a rubber band, for example, the
recovery to its equilibrium length is not, as one might think, exponential:
the relaxation is slow, in many cases logarithmic, and can still be observed after many hours. The form of the relaxation also depends on the
duration of the stretching, the `waiting-time'. This ubiquitous phenomenon is
called aging, and is abundant both in natural and technological applications.
Here, we suggest a general mechanism for slow relaxations and
aging, which predicts logarithmic relaxations, and a particular aging dependence
on the waiting-time. We demonstrate the generality of the approach by comparing our predictions to
experimental data on a diverse range of physical phenomena, from conductance in
granular metals, to disordered insulators, and dirty semiconductors, to the
low temperature dielectric properties of glasses.

\end{abstract}
\keywords{Aging | Glasses| Slow relaxations}
\
 \maketitle



\begin{figure*}[b!]
\includegraphics[width=0.8\textwidth]{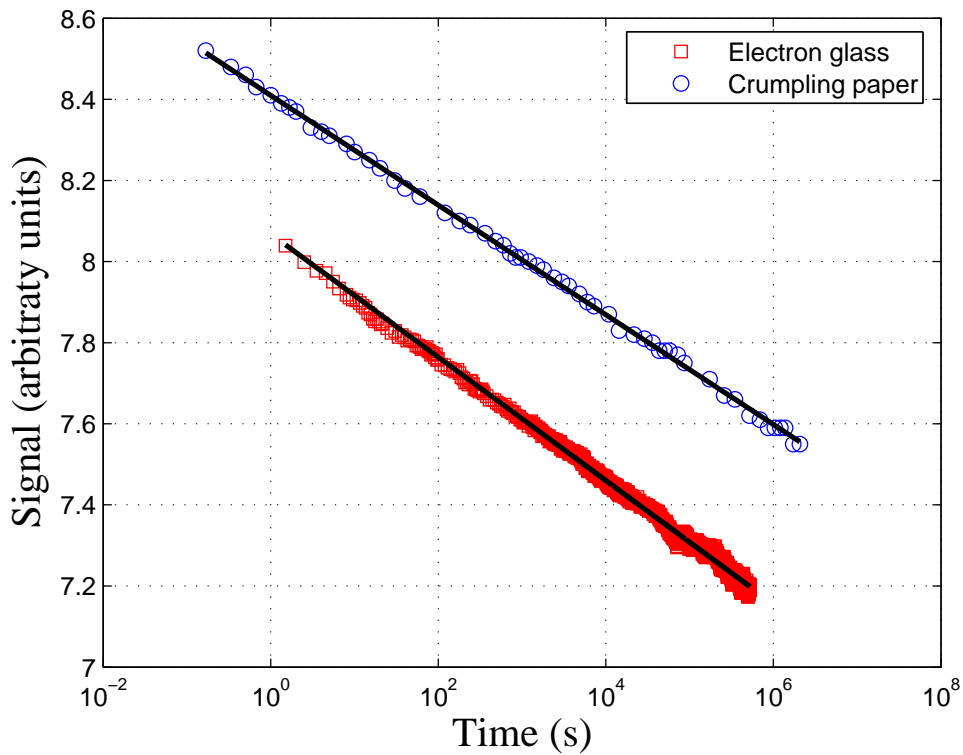}
\caption {Experimental results showing a logarithmic relaxation in the
electron glass indium oxide, where conductance is measured, and in a system of crumpling mylar, where the height is measured, after a sudden change in the experimental conditions. As seen in the graph, the logarithmic change in the physical observable can be measured from times of order of seconds or less to several days \cite{zvi2, crumpling}. Similar logarithmic relaxations, observed over many decades in time, occur in numerous physical systems, ranging from currents in superconductors to frictional systems. Data courtesy of Z. Ovadyahu and S. Nagel.} \label {log_plot}
\end{figure*}

Physicists often take for granted that systems relax exponentially. Indeed, when a capacitor discharges, it will discharge exponentially, with a rate independent of the time it has been charged for. However, the relaxation of many systems in nature is far from exponential, as was noticed already in the 19th century by Weber \cite{weber}. In many cases, the relaxation
is logarithmic: such relaxations have been experimentally observed in the decay of current in
superconductors \cite{creep_super}, current relaxation in MOSFET devices
\cite{mosfet_log}, mechanical relaxation of plant roots \cite{maize},
volume relaxation of crumpling paper \cite{crumpling} and frictional
strength \cite{jay}, to name but a few. Fig. \ref{log_plot} shows experimental data for electron glasses and of crumpling a thin sheet, that are governed by extremely different physical processes, yet they display identical relaxation behavior, which is logarithmic over a strikingly broad time window.

In these systems, in contrast to the capacitor example, the
relaxation \emph{does} depend on the time the system has been perturbed for -- in the scientific jargon, this is referred to as `aging'. In fact, slow
relaxations and aging are amongst the most distinct features of
glasses, whose understanding presents an important problem in
contemporary condensed matter physics. Much experimental and theoretical attention has been devoted to aging in the past decades, in a variety of fields, such as spin-glass \cite{aging_spinglass}, colloids \cite{aging_colloidal}, vortices in superconductors \cite{du} and many others \cite{aging_models}.

Here we study a generic model for aging, and discuss several mechanisms yielding a broad distribution of relaxation rates. We demonstrate the generality of the model on four different experimental systems, measuring the  dependence of the relaxation both on time $t$ and on the `waiting-time' $t_w$,
during which an external perturbation has been applied. We show how the following form of relaxations transpires: \be
S(t,t_w) \propto \log(1+t_w/t) = \left\{
\begin{array}{ll}
\log(t_w/t) & \text{for } t \ll t_w,\\
t_w/t       & \text{for } t \gg t_w.

\end{array} \right.
\label {aging} \ee
where $S$ is the physical observable.
 This means that the initial relaxation, at times short compared
with $t_w$, is logarithmic, while at long times compared to $t_w$ it
falls off as the inverse of the time -- a power-law decay, much
slower than exponential or stretched exponential decay \cite{bertin}.

\begin{figure*}[t]
\begin{center}
$\begin{array}{c@{\hspace{1in}}c}
\mbox{\bf (a)}
\epsfxsize=0.75\textwidth
\epsffile{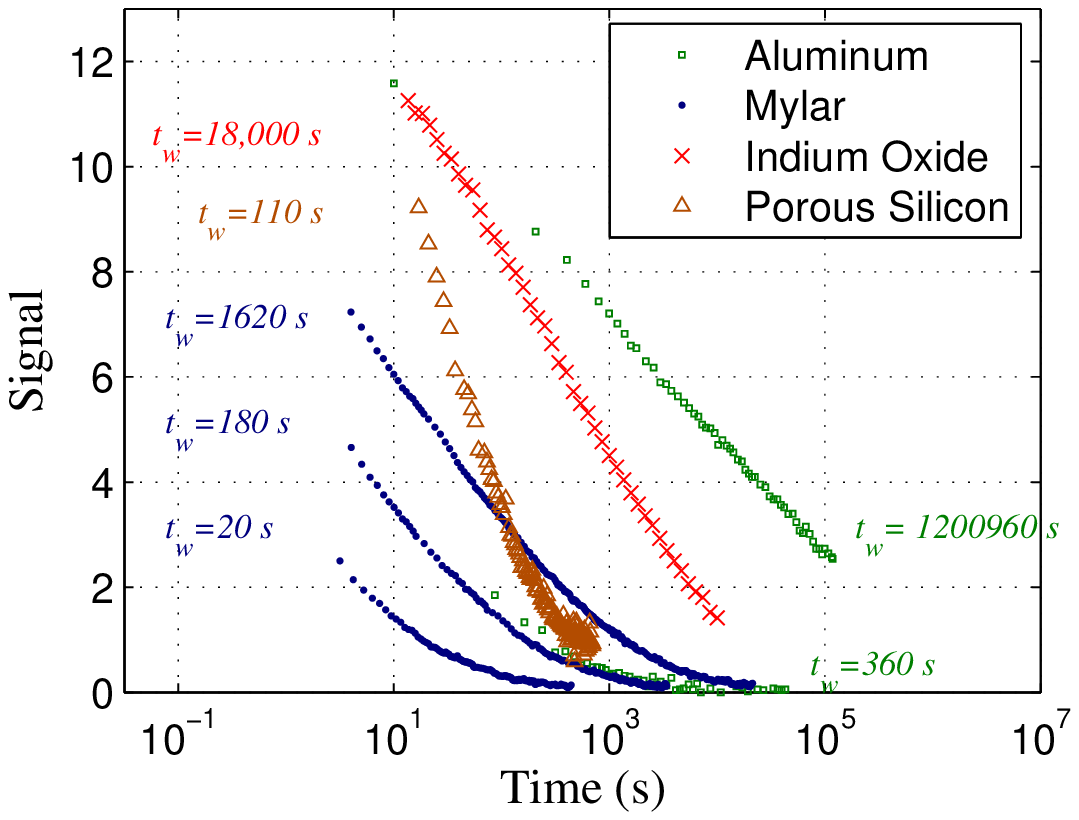}
\\  \mbox{\bf (b)}

	\epsfxsize=0.75\textwidth
	\epsffile{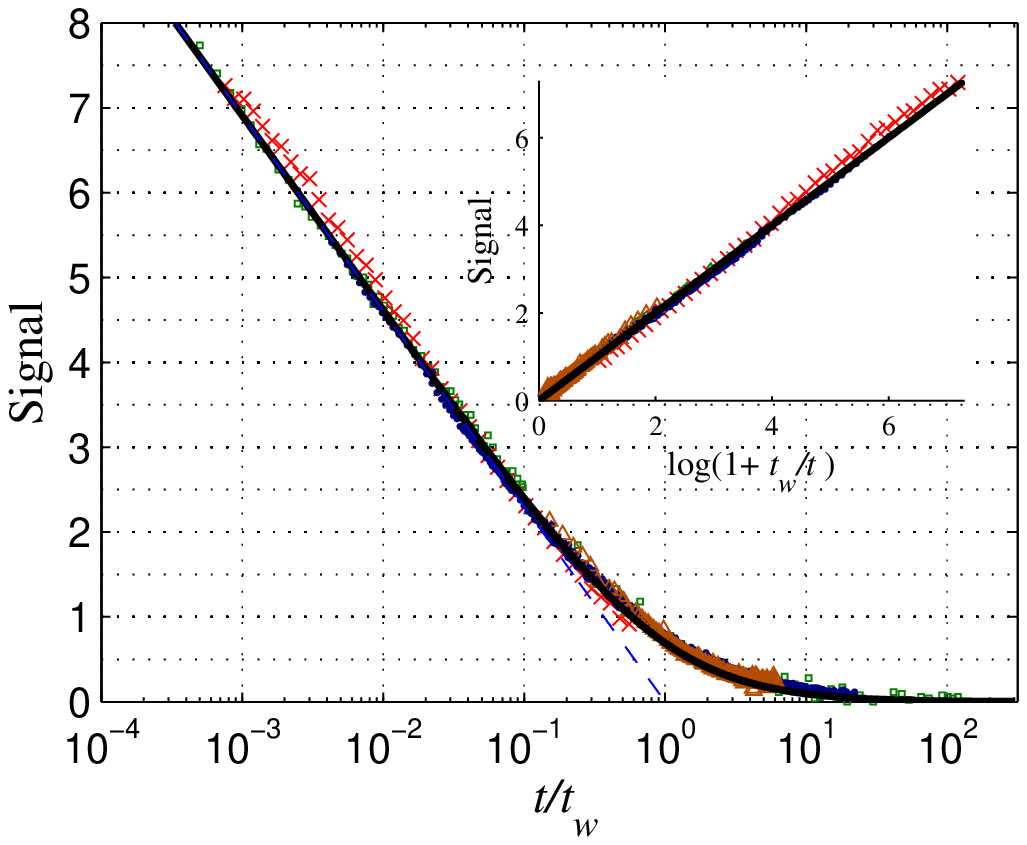}
\\

\end{array}$
\end{center}
\caption{(a) Results of aging experiments for four different
systems measuring different physical observables. Experimental
parameters can be found in Table \ref{exp_table}. The \emph{x} axis
denotes the time (on logarithmic scale, spanning five decades in
time, from seconds to days), and the \emph{y} axis denotes the
signal (with different units for each data set).
(b) The \emph{x} coordinate of each data set is scaled
according to the known waiting-time $t_w$, and the \emph{y}
coordinate is scaled such that the signal at $t=t_w$ is $\log(2)$
(for convenience). The data collapses onto a single curve, which is
compared to the theoretical prediction of Eq. (\ref{aging}). The
inset shows the same data has indeed linear dependence when the
\emph{x} axis is defined according to Eq. (\ref{aging}).}
\label {aging_fig} \end{figure*}

Fig. \ref{aging_fig} shows the excellent agreement between this prediction and experimental results
for four different systems, measuring various physical observables:
conductance relaxation in the electron glasses indium oxide and
granular aluminum
\cite{zvi1,zvi2,zvi3,zvi4,amir_glass,grenet1,grenet2,grenet3,amir_aging, amir_review},
relaxation of the dielectric constant in the plastic mylar
\cite{ludwig1,ludwig2,ludwig3} and conductance relaxation in room temperature
porous silicon \cite{borini_prb2007,borini_2011}. The experiments also markedly differ in the involved
timescales. Details of the experiments are given in Table~1.

~\begin{table*}[h!b!p!]
\caption{\textbf{Details of experiments}}
\begin{tabular}{lllll}
\hline
System & Measured variable & Units & $t_w$ (s) & Taken from\\
\hline
Aluminum & Conductance ($\sigma$) & $0.02 \delta \sigma /\sigma$  & 1200960, 360 & Grenet et al., Ref. \cite{grenet3} (2007). \\ 
Mylar & Dielectric constant ($\epsilon$) &  $10^{-6} \delta \epsilon /\epsilon $& 18000& Ludwig et al., Ref. \cite{ludwig1} (2003).\\ %
Indium Oxide & Conductance ($\sigma$)&  $0.01 \delta \sigma /\sigma$ & 20, 180, 1620& Ovadyahu, Ref. \cite {zvi4} (2006).  \\ %
Porous Silicon & Conductance ($\sigma$)& $0.02 \delta \sigma /\sigma$  &110& S. Borini, private comm. (2010). \\
\end{tabular}
\label{exp_table}
\end{table*}

In the following, we describe the aging protocol used in the experiments,
and introduce a model that predicts Eq. (\ref{aging}). We explain how one can understand the slow relaxations in terms of an underlying distribution of relaxation rates of a particular form:

\be  P(\lambda) \sim 1/\lambda \label {1_lambda},\ee

which we shall show can emerge due to various, physically distinct, mechanisms: thermal activation, quantum mechanical tunneling, or through a third mechanism relying on a multiplicative process. We then proceed to describe the connection to the ubiquitous 1/\emph{f} noise encountered in many systems, as well as the possible role of the distribution described in Eq. (\ref{1_lambda}) in other intriguing phenomena such as Benford's law \cite{benford}.

\section {The experimental protocol}

\begin{figure*}[b!]
\includegraphics[width=0.8\textwidth]{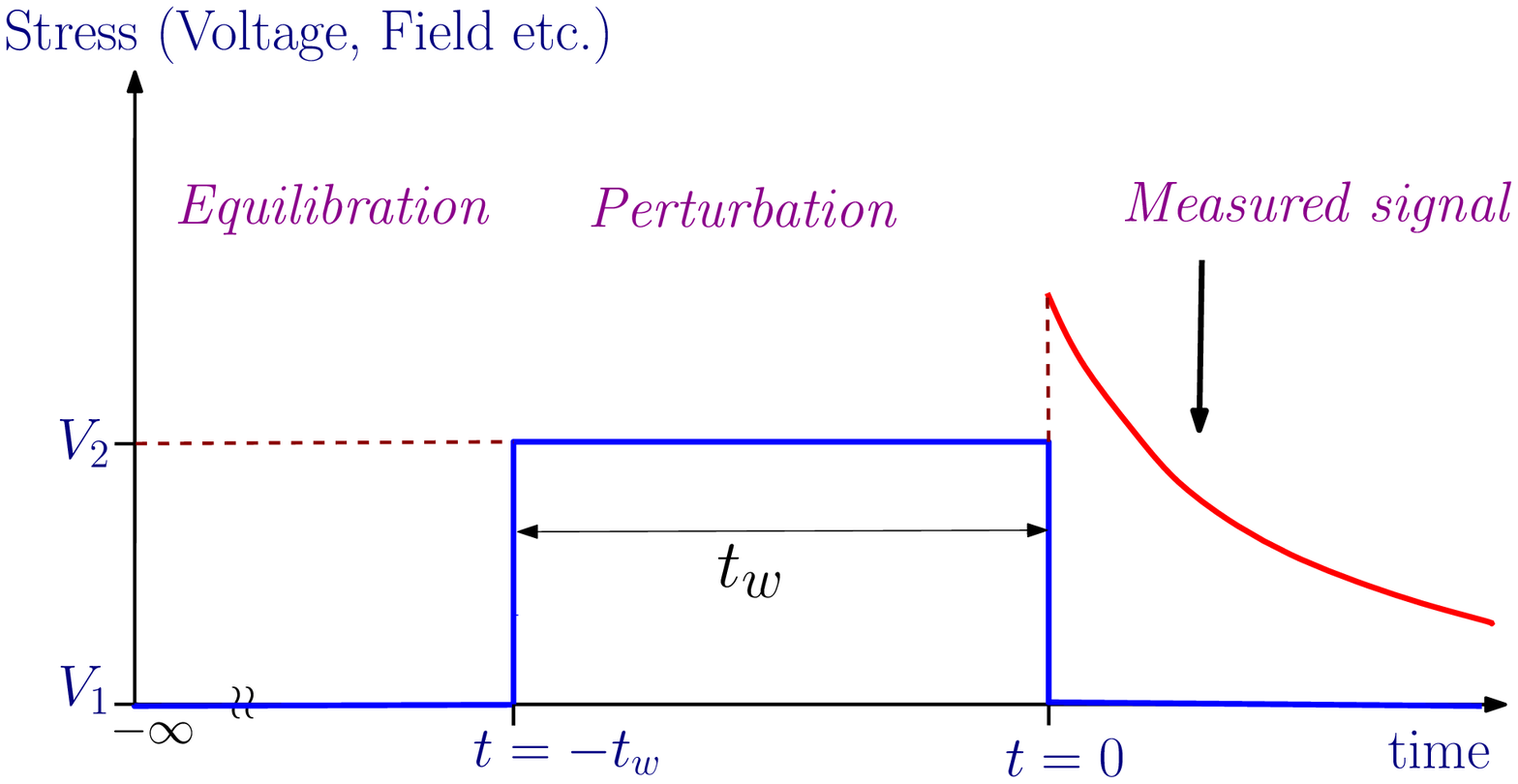} 
\caption {Schematic description of the different stages of the
aging protocol. At time $t=-t_w$ a perturbation is applied to the
system, which is turned off at time $t=0$. We will be interested in
the form of the relaxation of the physical observable in the last
stage.} \label {protocol}
\end{figure*}

The aging protocol is illustrated in Fig. \ref{protocol}. Its first step consists of letting the system attempt to equilibrate for a relatively long time (typically of the order of hours or days).
Next, one perturbs the system, in a way which depends on the
experimental system: for indium oxide and aluminum this is done by
changing the voltage of a capacitively coupled gate, for the mylar
sample it is done by putting the system in a perpendicular electric
field, while for porous silicon a large bias voltage is applied.
The perturbation is now maintained for a time $t_w$. After it is
switched off, the physical observable is continuously monitored, as
it relaxes. The longer $t_w$ is, the slower the resulting
relaxation.

\section {The model}

Having laid out a concrete experimental protocol we are now well positioned to describe a generic model which will yield the form of aging described by Eq. (\ref{aging}). The model will involve
two ingredients: First, the understanding that a broad distribution
of relaxation rates $\lambda$ occurs whose \emph{logarithm} is approximately uniformly
distributed over some broad range, as described by Eq. (\ref{1_lambda}).
The second ingredient involves understanding what this relaxation
rate distribution implies on the relaxations, within the aging
protocol.

The model assumes that the measured physical observable (conductance, dielectric constant etc.) is affected by an ensemble of modes, which are \emph{independent} and contribute to the observable in a definite way (for example, a relaxation of any of the modes will cause the conduction to \emph{decrease}). For each particular system under study, understanding the microscopic source of these modes is a quite different sort of question one may ask, and is outside the scope discussed here \cite{eglass_modes}. In general, these assumptions would make sense for a physical observable which depends on the configuration of the whole system (e.g: conductance, volume), and not some local probe (e.g: current measured in an STM tip). Also, we would always be assuming that a large number of these modes contribute, so that we can take the continuous limit, and discuss probability distributions. Presumably, the model would fail for a sufficiently small sample (although, in the case of electron glasses the logarithmic relaxations were experimentally observed even for micron sized samples \cite{zvi_mesoscopic}). Each of the modes would relax exponentially to its equilibrium, but with a different relaxation rate. To support these assumptions, one may think of the system as being formally characterized by a state vector $\vec{v}$, containing all the relevant information determining the physical observable. Perturbing the system weakly near its equilibrium, one can always linearize the equations of motion, and obtain an equation $\frac{d \delta \vec{v}}{dt}  =A \delta \vec{v}$, with $\delta \vec{v} \equiv \vec{v}(t)-\vec{v}_{equilibrium}$, and $A$ a matrix, independent of time. The real part of the eigenvalues of $A$ must be negative, in order for the equilibrium to be stable, and they have the physical meaning of relaxation rates: As is seen by solving the linear equation, each eigenmode relaxes exponentially to zero, independently of the other eigenmodes. In certain cases \cite{amir_glass}, the form of $A$ can be worked out explicitly.

We shall now explain three different mechanisms which lead to an abundance of slowly relaxing modes of the system, described mathematically by Eq. (\ref{1_lambda}). Using our assumption that the modes contribute positively and uniformly to the measured physical observable, the superposition of these modes will yield (in a certain time window, the conditions of which we will discuss) the logarithmic relaxations described earlier.

\subsection {Thermal activation}

A diversity of physical processes are governed by thermal activation, which is perhaps the simplest physical mechanism which can give rise to Eq. (\ref{1_lambda}), as has been known for long \cite{bernamont,VanDerZiel,bouchaud}. We should have in mind a rugged energy landscape characterizing the system, with many local minima. At a given time, we can denote by $\vec{p}$ the vector of probabilities for the system to reside in each of the minima. Clearly, for a system governed by stochastic dynamics, the probability vector would obey the same linear equation mentioned earlier, namely, $\frac{d\vec{p}}{dt}=A\vec{p}$ (\emph{i.e}.,  we have defined a Markov process).
The relaxation modes in this case approximately correspond to crossing one of the energetic barriers connecting two of the local minima.
Here, the rate $\lambda$ of a given process is given by the Arrhenius formula, namely, it is exponential in the energetic barrier $U$, namely $\lambda
\propto e^{-U/kT}$, which the system has to cross in order to reduced
its energy. We will associate each mode with one such transition (across an energetic barrier $U$), and assume that the size of these barriers is distributed smoothly over a
certain range of energies. We can now readily calculate the resulting relaxation
rate distribution:  \be \rm{P}(\lambda) = \frac{\rm{P}
(\emph{U})}{|d\lambda/d\emph{U}|} \sim T/\lambda, \label{activation}
\ee where we have taken $\rm{P}(\emph{U})$ as approximately
constant. For isothermal processes, the temperature dependence
entering the proportionality constant does not play a role in the
aging behavior. From this formula we can also deduce the smallest
and largest rates the system supports (corresponding to the fastest
and slowest times): these are $\lambda_{\rm{min}} \propto
e^{\frac{-U_{\rm{max}}}{kT}}$ and $\lambda_{\rm{max}} \propto
e^{\frac{-U_{\rm{min}}}{kT}}$, related to the extremal barrier heights. Note
that due to the exponential dependance on $U$, even a small range of energy barriers can result in a broad
range of relaxation rates. The above mechanism is essentially the same leading to 1/\emph{f} noise \cite{dutta,Weissman}, and is reminiscent of Bouchaud's trap model \cite{bouchaud}.

It should be emphasized that we have assumed here that the energy barriers vary sufficiently slowly, and
therefore their variation within the energy interval $[U_{\rm{min}},
U_{\rm{max}}]$ can be neglected. This simple picture turns out to be
extremely successful when applied to recent experiments on porous
silicon performed around room temperature. At such a high
temperature, thermal effects are expected to be dominant over
quantum effects, and indeed the maximal timescale
$\lambda_{\rm{max}}$ was found experimentally to be very sensitive
to temperature, as expected from the above formula
\cite{borini_2011}.

However, it is experimentally found that in various systems the relaxations are insensitive to temperature \cite{grenet3}, which necessitates a different mechanism. Quantum Tunneling (QM) is the second mechanism which yields Eq.
(\ref{1_lambda}).

\subsection {Quantum mechanical tunneling}

Let us keep in mind the picture described earlier for the rugged energy landscape, but now assume that we are at low enough temperatures such that thermal activation across the barriers is prohibited. The system will be able to quantum mechanically tunnel through the barriers, paying a penalty which is typically exponentially suppressed with the distance. Thus, for this process as well, the rate $\lambda$ depends
exponentially on a smoothly distributed variable, which in this instance is
the distance: $\lambda \sim e^{-2r/\xi}$, with $\xi$ the
localization length of the wavefunctions, and $r$ the spatial distance between the two points. In a recent work
\cite{amir_expmat}, the distribution of relaxation rates was
calculated for this case, taking into account the correlations that exist (the distances are not independent in this case). It was found that one still obtains the $1/\lambda$ distribution, albeit with small but interesting
logarithmic corrections. Here, the role played by the temperature
$T$ in Eq. (\ref{activation}) is played by the localization length
$\xi$. Related considerations for a varying height of the barrier
through which the tunneling occurs are given in Ref. \cite{harrison}.

In both cases, of thermal activation and of quantum mechanical
tunneling, we would like the barrier distribution to be broad in energy or
real space, namely, it should be large compared to the energy $kT$ or the
localization length $\xi$, respectively, in order to achieve a broad
range of relaxation rates.

So far we discussed two different natural ways which lead to it, namely, thermal activation \cite{bernamont,VanDerZiel,bouchaud} and quantum tunneling \cite{amir_expmat}. The exponential nature of these processes is the key ingredient in obtaining Eq. (\ref{1_lambda}). Both mechanisms, however, are inadequate to describe the logarithmic relaxation in crumpling paper, for example \cite{crumpling}. We shall now present another mechanism, which does not rely on a variable being exponential, but rather, on the central-limit-theorem. This suggests the mechanism should
be broadly applicable. We will show how the interplay of many random
processes can under general conditions lead to a log-normal
distribution, which is well approximated over a broad range by Eq.
(\ref{1_lambda}).

\subsection {Multiplicative processes}

In many physical examples, an observable depends on
the \emph{product} of many approximately independent variables, which is referred to as a multiplicative process. Understanding the importance of such processes in nature dates back (at least) to Shockley \cite{shockley}, who discussed the connection of multiplicative processes to log-normal distribution which we shall also utilize here. See \cite{economic} for a strongly related discussion in the context of 1/\emph{f} noise.

 An example of a multiplicative process is the transmission of a particle through a one-dimensional disordered wire: if we divide the wire into a large number of segments, it is
known that the average transmission is the product of the individual
transmissions through each segment \cite{imry_mesoscopics}. Fig.
4 demonstrates pictorially another such example,
namely, how for electron glasses relaxation can occur via the
simultaneous tunneling of various electrons, which is also
approximately a multiplicative process. If we assume that
the relaxation rate $\lambda$ is a product of many independent
variables $x_i$, we can readily calculate the resulting distribution
of relaxation rates. Since $\lambda= \prod_i{x_i}$, we have, using
the central limit theorem: \be \rm{P} \left(\log(\lambda)\right) \rightarrow
e^{-{[\log(\lambda)-\mu]/\Delta}^2}. \ee By changing back to the
variable $\lambda$, we find that it follows a log-normal
distribution: \be P(\lambda) \sim e^{-[\log(\lambda t_0)/\Delta]^2} /\lambda \label{lognormal},\ee where $t_0$ is a constant with the dimensions of time. Far from the
tails of the distribution, namely, when $|\log(\lambda t_0)| \ll \Delta$ the distribution reduces to
Eq.~(\ref{1_lambda}). Remarkably, in the crumpling paper example, the distribution of the lengths of the segments was measured directly, and shown to follow a log-normal distribution \cite{papergeometry}. The compatibility of log-normal distribution and logarithmic relaxations fits well with the theoretical framework we suggest.

\begin{figure*}[h]
\begin{center}
$\begin{array}{c@{\hspace{0.5in}}c} \multicolumn{1}{l}{\mbox{\bf
(a)}} &
    \multicolumn{1}{l}{\mbox{\bf (b) }} \\ [-0.0cm]
\epsfxsize=2.8in \epsffile{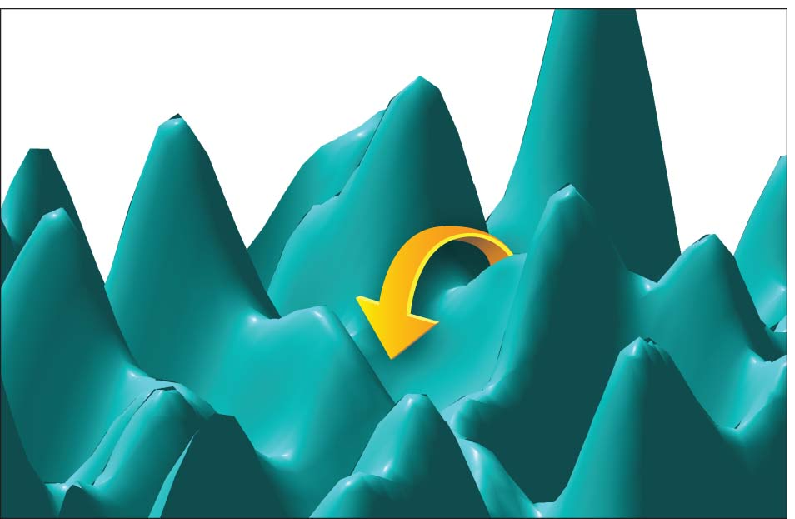} &
    \epsfxsize=2.8in
    \epsffile{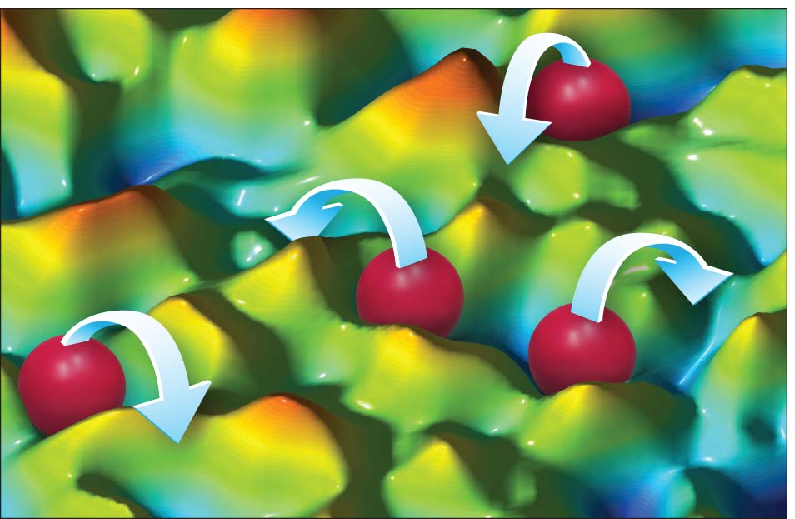} \\ [0.2 cm]
\mbox{\bf  } & \mbox{\bf}
\end{array}$
\end{center}
\caption{Pictorial demonstration of different physical mechanisms leading to a broad distribution of relaxation rates. (a) The energy landscape of various complex systems, including glasses, contains many minima. The energetic barriers separating them are smoothly distributed over a certain range. In order for the system to relax its energy, it must cross these barriers by thermal activation or by quantum mechanical tunneling: both are exponential in the barrier, which lead to a rate distribution described by $P(\lambda) \sim 1/\lambda$, as we discuss in detail. (b) Many particle transitions in an electron glass are an example of a multiplicative process: In many electronic configurations, moving any single electron in the system to one of the vacant sites will result in higher energy, and therefore these processes will not occur at low enough temperatures. However, changing the position of a larger
number of electrons can result in a lower energy. The rates of this process can be approximately written as a product of the rates of the single particle processes involved, leading to a rate distribution described by a log-normal distribution.}
\label{multiple}
\end{figure*}

%

\section {Derivation of the aging formula}

We shall now discuss the implications of this distribution on aging experiments, showing it leads to aging of a particular form, called `full' aging, in a particular limit. This was done in the context of relaxation in electron
glasses in \cite{amir_aging}. The derivation we will present, however, does not rely on any peculiar
properties of this specific system, and as such will be broadly
applicable for all physical systems which follow the $1/\lambda$
distribution.

Let us assume that the system supports $i=1\dots N \gg 1$ relaxation
modes, with corresponding relaxation rates $\lambda_i$, each of
which contributes an amount $X_0$ to the physical observable
(\emph{e.g.}: conductance or dielectric constant). Before going
to the more involved aging experiment, let us consider the case
where we excite all of these modes by some uniform amplitude. The
physical observable $X$ measured a time $t$ after the perturbation,
would read $X(t) = X_0 \sum_i e^{-\lambda_i t}$, which in the
continuous limit goes to:

\be X(t)= X_0 \int_{\lambda_{\rm{min}}}^{\lambda_{\rm{max}}} d\lambda P(\lambda) e^{-\lambda t}, \ee where
$\rm{P}(\lambda)$ is the distribution of relaxation rates. In the
case where $\rm{P}(\lambda) \sim 1/\lambda$, introduced earlier, with $\lambda_{\rm{min}}$ and $\lambda_{\rm{max}}$ the lower and upper cutoffs, we obtain the difference of two exponential integral function \cite{pollak, amir_aging}:

\be X(t)= X_0[E_1(\lambda_{\rm{min}}t)-E_1(\lambda_{\rm{max}}t)].\ee

For the case where the involved times are much smaller than the
reciprocal lower cutoff $\lambda_{\rm{min}}$, and much larger than the reciprocal upper
cutoff, the equation reduces to a simpler form:

\be X(t)= X_0 [-\gamma_E - \log(\lambda_{\rm{min}}t)   ]. \ee Thus,
we expect a logarithmic relaxation, which is indeed experimentally
observed in a large variety of systems, as discussed earlier. Fig.
1 shows such a logarithmic relaxation, measured in an
indium oxide sample.

Going on to the aging protocol, we shall assume that the
perturbation is small enough such that the rates of the relaxing
modes are indifferent to it. Nonetheless, upon the application of
the perturbation, the fixed point to which the system attempts to
relax to (which does not have to be the true equilibrium, but can be
a metastable state), is different when the perturbation is applied.
Therefore during the second stage of the experiment (see Fig.
4), the system relaxes towards the new metastable
state, which means that the relaxation modes are excited with
respect to the \emph{original} metastable state: relaxation to the new metastable state implies excitation with respect to the old one. The closer we got to the new minimum, the further we are from the initial one.

Let us illustrate this for the example of a single relaxation mode: in this case the relaxation is exponential, and therefore at the moment when the perturbation is switched off the distance from the new metastable state is proportional to $e^{-\lambda t_w}$. At this moment the distance from the original metastable state is $1-e^{-\lambda t_w}$: indeed, for $t_w=0$ nothing happens, while for $t_w \rightarrow \infty$ the largest possible excitation occurs. A time $t$ later, the amplitude of the relaxation mode, which decays exponentially, would therefore be $(1-e^{-\lambda t_w})e^{-\lambda t}.$ Generalizing this for the case of many relaxation modes, we find that a time $t$ after the perturbation has been switched off, the physical observable is given by:

\be X(t)= X_0 \int_{\lambda_{\rm{min}}}^{\lambda_{\rm{max}}} d\lambda P(\lambda) (1-e^{-\lambda
t_w})e^{-\lambda t}. \label{full_form} \ee

This can be written, as before, in terms of exponential integral function.
Assuming that we are in the regime of intermediate asymptotics, where the experimental timescales are much larger than the reciprocal upper cutoff and much smaller than the reciprocal lower cutoff, we obtain the difference of two logarithms:

\be X(t)/X_0= \log[\lambda_{\rm{min}}(t+t_w)]-\log[\lambda_{\rm{min}}t], \label{aging2} \ee

leading to Eq. (\ref{aging}).

It should be noted that the regime where the experimental time is comparable to $1/\lambda_{\rm{min}}$ can also be reached, and it was shown that Eq. (\ref{full_form}) accounts of the aging behavior in porous silicon also in the case where significant deviations from the full aging regime were observed \cite{borini_2011}. In other words, the model predicts full aging only in the asymptotic regime, and can also account for the deviations from full aging. Related system dependent cutoffs were also discussed in the context of spin-glass \cite{bouchaud_cutoff}.


\section {Connection to $1/f$ noise}

The broad underlying distribution of relaxation rates which played a crucial role in determining the slow relaxations, is also central for understanding low-frequency noise in many systems. A variety of physical, biological and financial models show a universal form of low-frequency noise \cite{dutta, Weissman, heartbeat, economic}, with a power-spectrum scaling as 1/\emph{f}. This ubiquitous form of noise is deeply related to the logarithmic relaxation which we study here, the underlying principle being a roughly uniform distribution of
effective barriers. In \cite {amir_noise},  a relation between the two physical phenomena was made, based on a theory devised by Onsager nearly a century ago: the connection is made through the Onsager's regression principle, stating that the relaxation
of a system close to its equilibrium is related to the spectrum of the fluctuations of the system
around the equilibrium \cite{onsager}. Each mode with a rate $\lambda$, generates a Lorentzian noise spectrum \cite{bernamont,VanDerZiel}:
\be I(f) \propto \frac{\lambda}{f^2+\lambda^2}. \ee Summing over many modes using the distribution of Eq. (\ref{1_lambda}), yields 1/\emph{f} noise. Linking these two seemingly unrelated universal behaviors, logarithmic relaxations and 1/\emph{f} noise, seems to us both conceptually appealing and of practical importance. For one, it means that the different physical mechanisms we suggested to yield the broad distribution of relaxation rates, would also imply 1/\emph{f} noise.

\section {Other appearances of the $1/\lambda$ distribution in nature}
In fact, the $P(\lambda) \sim 1/\lambda$ distribution which plays a pivotal role in determining the aging,
logarithmic relaxations and 1/f noise, appears also in completely different contexts.
The Gutenberg-Richter law \cite{Richter}, for example, states that the distribution of
the magnitude of earthquake is a power-law. The exponent is experimentally found, in various
methods of analysis, to be close to one \cite{quake}. Another striking example lies in the so-called
`first-digit problem'.

Towards the end of the 19'th century, the astronomer Simon Newcomb noticed, while looking
at his logarithm table, that numbers starting with 1 are looked up far more often than higher
digits \cite{newcomb}. Half a century later, the physicist Frank Benford rediscovered the phenomenon, and
asked himself the following question: what is the distribution of the leading digits of numbers
encountered in a certain scenario? \cite{benford}. Remarkably, in hugely differing data sets such as those
found in tax returns, tables of physical constants, birth rates and many others, the relative occupance
of each digit follows a universal (and nonuniform) distribution: $P(d) = \log(1 + 1/d)$,
where \emph{d} is the digit. This is known as Benford's law, and is robust enough to be used to detect frauds in tax returns \cite {tax}. It can be explained in a simple way, if we assume that the distribution of $x$, the variable measured, follows approximately $P(x) \sim 1/x$,
over a large window. Clearly, the probability that the first digit is 1 is proportional to $\int_1^2 1/x \,dx + \int_{10}^{20} 1/x \, dx+...= N \log [2]$, where N is the number of decades spanned by the distribution (assumed to be a large number). Similarly for the digit \emph{d}, one obtains $N \log[(d+1)/d]$, yielding Benford's law. Thus, there is an intimate relation between the $1/x$¸ distribution leading
to logarithmic, slow relaxations, and Benford's law, stating universal statistics in the
first digit problem. It remains a challenge to find the unifying mechanisms between these various
observations.
%

\section{The extent of applicability of the theory}
So far, we have discussed three different physical mechanisms leading to the same aging behavior. There are, of course, other forms of slow relaxations in nature, which are also commonly observed, and many fascinating physical systems which do not fit into the framework outlined here. Two important examples are the aging behavior of spin-glass systems, which appears to be more complex \cite{horacio, vincent, spin_glasses} and is not described by this model, and the behavior of molecular and colloidal glasses, which have received much attention in recent years \cite{berthier}. As indicated by Refs. \cite{langer1}, \cite{langer2}, the model proposed here is not consistent with the relaxations observed in these systems.
Another generic form of slow relaxations in nature is stretched exponential relaxations $e^{-(t/\tau)^\beta}$, which has been experimentally observed in various systems \cite{weber, kohlrausch, phillips, du, hebard}, and for which several theoretical models have been proposed \cite{palmer,Huber,moore,Sturman}. For small $\beta$, this form is similar to Eq. (\ref{aging}), but it is still possible to distinguish the two, by analyzing the short-time behavior, where a power-law markedly differs from a logarithm.
 It would be interesting to make a better classification of these two universality classes (and possibly others), and to determine the extent and limitations of the applicability of each of them, but this task is beyond the scope of this work.

\section{Conclusions}
To summarize, in this paper we discussed a generic model for slow
relaxations and aging, whose signature is a distinct crossover from a
logarithm to a power-law, with no fitting parameters other than the overall scaling. Theoretically, we have shown how various different physical mechanisms give rise to a broad distribution of relaxation rates, of a particular form, and analyzed the resulting aging behavior. The data from various experiments was shown to agree well with this
prediction, over many decades in time. The experiments measure different physical observables, in a variety of systems and temperatures. This form of aging is fundamentally connected to other phenomena which are commonly observed in nature, such as 1/\emph{f} noise.
Part of the beauty of physics lies in the surprising connections it offers, between different fields and phenomena. We showed that one could understand on equal footing the aging of quantum tunneling in electron glasses
\cite{amir_aging} and the mechanical relaxations of plant roots \cite{maize}, and that both are connected to the 1/\emph{f} noise electrical engineers are well familiar with. As such, we believe the model presented can serve as a paradigm for slow relaxations and aging for a broad class of systems.





\begin{acknowledgments}
We thank S. Borini, J. Delayhe , T. Grenet,
S. Nagel and Z. Ovadyahu for important discussions and for their experimental data, and J. Langer, A. Hebard and J. P. Bouchaud for useful comments. This work was supported by a BMBF DIP grant as
well as by ISF and BSF grants and the Center of Excellence Program. The research of YI was supported by a Humboldt extension grant. \end{acknowledgments}

%
%




\end{document}